\documentclass[a4paper,12pt]{article}
\usepackage{amsmath}
\usepackage{graphicx}
\usepackage{cite}
\usepackage{setspace}
\usepackage{rotating}

\title{Energy levels of an anharmonic oscillator in both weak and strong coupling 
limit using convergency of Morse-Feshbach non-linear perturbation series}
\author{Biswanath Rath$^1$, Pramoda Kumar Samal$^2$, \\ Radhanath Mishra$^2$ and Basudeb Sahu$^1$ }
\date{}

\begin{document}
\maketitle
\begin{center}
{$^{1}$Department of Physics, North Orissa University, Baripada, 757003, India \\ }
{$^2$Department of Physics, Utkal University, Bhubaneswar, 751004, India\\}
\end{center}
 
\begin{abstract}
We make an extensive rigorous study on convergent behaviour of Morse-Feshbach 
nonlinear perturbation series (MFNPS) to find out energy levels of the
anharmonic oscillator (AHO) in both weak and strong coupling limit. 
We develop a new method of multi step optimal splitting in order to 
get convergency in MFNPS for ground state of AHO and found that
two step optimal splitting is sufficient to provide convergency in MFNPS. Unlike 
the ground state the optimal splitting parameters for excited states is modified 
according to their dependency on state in order to achieve convergency in MFNPS.
\end{abstract}


\section{Introduction}

In quantum mechanics, there are only few 
problems whose Hamiltonians can be solved exactly. However for any
arbitrary Hamiltonian one has to use some sort of approximations [1].
Perturbation theory is one of the few principal methods of approximation for
solution to eigenvalue problems in quantum mechanics [1-10]. In perturbation 
theory one needs to solve the Hamiltonian of the type
\begin{equation}
H = H_{0} + \lambda H_{1}
\label{eqn1}
\end{equation}
where $H_{0}$ is the unperturbed Hamiltonian, which can be solved exactly 
 and $\lambda H_{1}$ is the 
perturbation term. If $E$ corresponds to energy of the total Hamiltonian $H$
and $E_n^0$ corresponds to unperturbed Hamiltonian $H_{0}$, then one can 
obtain $E$ in two different ways as 
\begin{equation}
E =f(E_n^0, \, \lambda)
\label{eqn2}
\end{equation}
or
\begin{equation}
E =f(E,\,E_n^0 ,\, \lambda ) 
\label{eqn3}
\end{equation}

Let us analyse the above two relations. From the relation in Eq (\ref{eqn2}) it is evident
that R.H.S is a function of known energy $E_n^0$ and coupling constant. This method of
calculation is known as constant function perturbation theory as seen from the 
literature [1-8]. Now analyse the the relation in Eq (\ref{eqn3}). In this case
both L.H.S and R.H.S are the function of unknown energy $E$. This type of perturbation 
series is known as nonlinear series. The most simplest nonlinear perturbation
series is the Morse-Feshbach nonlinear perturbation series (MFNPS) [9-10].
Further it is seen that almost all attemption of perturbation theory reflect Eq (\ref{eqn2}) 
with AHO as an example. So far there is very few work on nonlinear series reflecting 
Eq (\ref{eqn3}). Because of highly nonlinear in nature MFNPS is extremely difficult to solve 
analytically and needs high computational skill to solve it numerically in order to get 
higher order convergency. Even the most simplest form of nonlinear series is the MFNPS is 
in literature for more than fifty years in literature, only it has been applied only to 
the ground state of AHO [10]. The basic aim of this paper is to study the 
applicability of this series by calculating excited state energy considering AHO
as an example. In a more explicit version of the aim is to address the followings:

\begin{itemize}
 \item whether the convergency on ground state energy obtained previously [10] can be 
improved by introducing multi step optimal splitting parameter?

\item whether the  previous method applied for ground state will remain valid for the excited state energy calculation?

\item whether for excited state energy calculation one has to make some modification?

\end{itemize}

To arrive at a suitable conclusion to above questions we have to make a
 systematic study in the following sections.

\section{Morse-Feshbach nonlinear perturbation  series (MFNPS)}

In this section we write the MFNPS in a simplified language .
Let us  solve the unperturbed Hamiltonian 
\begin{equation}
H_{0}\, |n\rangle=E_{n}^{(0)}\,|n\rangle
\label{eqn4}
\end{equation}
where $|n\rangle$ is the unperturbed eigenfunction corresponding to energy eigenvalue 
$E_{n}^{(0)}$. According to MFNPS the energy $E$ for the perturbed Hamiltonian $H$ is given as
\begin{eqnarray}
E_{n} &=& E_{n}^{(0)} +\lambda \, \langle n|H_{1}|n \rangle + \lambda^{2}\, 
\sum_{m \neq n}\frac {\langle n|H_{1}|m \rangle \, \langle m|H_{1}|n\rangle}{(E_{n}-E_{m}^{(0)})} \nonumber \\
&+& \lambda^{3} \,\sum_{m,k \neq n} \frac {\langle n|H_{1}|m \rangle \, \langle m|H_{1}|k\rangle \, \langle k|H_{1}|n \rangle}{(E_{n}-E_{m}^{(0)})(E-E_{k}^{(0)}} \nonumber \\
&+& \lambda^{4} \, \sum_{m,k,p \neq n}\frac{\langle n|H_{1}|m\rangle \, \langle m|H_{1}|k\rangle\,\langle k|H_{1}|p\rangle \,\langle p|H_{1}|n\rangle}{(E_{n}-E_{m}^{(0)})(E_{n}-E_{k}^{(0)})(E_{n}-E_{p}^{(0)}) } \nonumber \\
&+&\lambda^{5} \, \sum_{m,k,p,q \neq n}\frac {\langle n|H_{1}|m\rangle\,\langle m|H_{1}|k\rangle\,\langle k|H_{1}|p\rangle \, \langle p|H_{1}|q\rangle\,\langle q|H_{1}|n\rangle}{(E_{n}-E_{m}^{(0)})(E_{n}-E_{k}^{(0)})(E_{n}-E_{p}^{(0)})(E_{n}-E_{q}^{(0)}) } \nonumber \\
&\cdots& + \,\, \lambda^{K} \sum_{m \cdots z \neq n}\frac {\langle n|H_{1}|m\rangle \cdots \, \langle z|H_{1}|n\rangle}{(E_{n}-E_{m}^{(0)})........(E_{n}-E_{z}^{(0)}) }
\label{eqn5}
\end{eqnarray}
Here $K$ is the order of perturbation in MFNPS.

\section{Higher order calculation on convergence test for groundstate using two step optimal splitting}

In this section we test the convergence of the ground state energy of the 
AHO, whose Hamiltonian is given as
\begin{equation}
H=\frac{ p^{2}}{2}+\frac{ x^{2}}{2}+\lambda\, x^{4}
\label{eqn6}
\end{equation}
using two step optimal splitting in higher order calculation.
The coordinates $x$ and momentum $p$ satisfy the relation
\begin{equation}
[x,p] = i\hbar
\label{eqn7}
\end{equation}
Now we rewrite the Hamiltonian in Eq (\ref{eqn6}) as 
\begin{equation}
H=\frac{ p^{2}}{2} + w^{2} \, \frac{ x^{2}}{2} + \lambda \, x^{4} - f(x)
\label{eqn8}
\end{equation}

where 
\begin{equation}
f(x)=\frac {6 \lambda x^{2}}{2w}
\label{eqn9}
\end{equation}
and 
\begin{equation}
w^{2}= 1 + \frac {6\lambda}{w}
\label{eqn10}
\end{equation}
Further we rewrite the Eq (\ref{eqn8}) as 
\begin{equation}
H=\frac{ p^{2}}{2}+ W_{1}^{2} \frac{ x^{2}}{2}+\lambda x^{4} - f(x) -  F_{1}(x)
\label{eqn11}
\end{equation}
where 
\begin{equation}
F_{1}(x)=\frac {6\lambda x^{2}}{2W_{1}}
\label{eqn12}
\end{equation}
and
\begin{equation}
W_{1}^{2}= w^{2} + \frac {6\lambda}{W_{1}}
\label{eqn13}
\end{equation} 
Now we rewrite the Hamiltonian as 
\begin{equation}
H = H_{D} +H_{N} 
\label{eqn14}
\end{equation} 
where $H_{D}$ is the diagonal term and considered as unperturbed Hamiltonian.
Similarly $H_{N}$ is the non-diagonal term and considered as perturbation term.
The explicit expression for $H_{D}$ in terms of creation operator $a^{\dagger}$ and 
annihilation operator $a$ is 
\begin{equation}
H_{D}= \frac{\left(W_1 + \frac{\textstyle{1}}{\textstyle{W_1}}\right) (2\,a^{\dagger}a +1)}{4} +
\lambda \, \,\frac{3+ 12 a^{\dagger}a + 6 (a^{\dagger})^2 a^2}{4W_1^2}
\label{eqn15}
\end{equation} 
Similarly the expression for $H_{N}$ in terms of creation operator $a^{\dagger}$ and 
annihilation operator $a$ is 
\begin{equation}
H_{N}= \lambda \, \left[\frac{a^4 + (a^{\dagger})^4 + 4 (a^{\dagger})^3 a + 4 a^{\dagger} a^3}{4W_1^2} -
\frac{3(a^2 + (a^{\dagger})^2)}{2 w W_1} \right ]
\label{eqn16}
\end{equation} 

The non zero expectation values of the diagonal Hamiltonian $H_D$ are given as 

\begin{equation}
\langle n |H_D| n \rangle =\left(\frac{2n+1}{4}\right)\left (W_1 + \frac{1}{W_1} \right) + \frac{3 \lambda}{4 W_1^2}(2n^2 + 2n +1)
\label{eqn17}
\end{equation}

The non zero expectation values of the non-diagonal Hamiltonian $H_N$ are given as 

\begin{equation}
\langle n |H_N| n+2 \rangle =\frac{ \lambda \sqrt{(n+1)(n+2)}}{W_1} \, \left(\frac{n}{W_1} - \frac{3}{2 w}\right)
\label{eqn18}
\end{equation}

\begin{equation}
\langle n |H_N| n+4 \rangle =\frac{ \lambda \sqrt{(n+1)(n+2)(n+3)(n+4)}}{4W_1^2}
\label{eqn19}
\end{equation}

Now replacing $\lambda H_{1}=H_{N}$ and $H_{0}=H_{D} $ in Eq (\ref{eqn1}, \ref{eqn4}, \ref{eqn5})
 we calculate the groundstate energy of the AHO up to 
$14$ order and the results are given in Table\ref{tab1}.

\begin{table}[h]
\centering
\begin{tabular}{ccccc}
\hline
\hline
& Order &  &  &  Convergent value $E_0$ with   \cr
& K     &  &  &  $w= 2.0$                \cr
&       &  &  &  $W_1=2.52510225481$     \cr
\hline
\hline
& 0 & & & 	0.847907429  \cr
& 1 & & & 	0.847907429  \cr
& 2 & & & 	0.8041081069829833  \cr	
& 3 & & & 	0.8039091999664616  \cr
& 4 & & & 	0.8037978280048715  \cr
& 5 & & & 	0.8037792238820139  \cr
& 6 & & & 	0.8037726920274627  \cr
& 7 & & & 	0.8037713351909546  \cr
& 8 & & & 	0.8037708048619517  \cr
& 9 & & & 	0.8037707015253137  \cr
& 10 & & &	0.8037706427848315  \cr
& 11 & & &	0.8037706383463737  \cr
& 12 & & &	0.8037706263550815  \cr
& 13 & & &	0.8037706301691552  \cr
& 14 & & &	0.8037706243946594  \cr
& 15 & & &	0.8037706287404748  \cr
\hline
\hline
\end{tabular}
\caption{ Convergent value for ground state energy $E_0$ of an AHO with $\lambda =1$ in  MFNPS using two-step optimal splitting parameters.}
\label{tab1}
\end{table}

\section{Higher order calculation on convergence test for groundstate using multi step optimal splitting}

\begin{table}[h]
\centering
\begin{tabular}{cccccccc}
\hline
\hline
  	&  &   Convergent value $E_0$     & &  Convergent value $E_0$      & &  Convergent value $E_0$     \cr
      	&  &   $w= 2.0$             & &  $w= 2.0$              & &  $w= 2.0$             \cr
 Order  &  &   $W_1=2.52510225481$  & &  $W_1=2.52510225481$   & &  $W_1=2.52510225481$  \cr
 K      &  &   $W=2.90538656129$    & &  $W_2=2.90538656129$   & &  $W_2=2.90538656129$  \cr
        &  &                        & &  $W_3=3.21090388686$   & &  $W_3=3.21090388686$  \cr
        &  &                        & &                        & &  $W_4=3.46974595594$  \cr
\hline
\hline
 0 & & 0.901242878		& &	0.953331243		& &	1.00178467		\cr
 1 & & 0.901242878		& &	0.953331243		& &	1.00178467		\cr
 2 & & 0.8022224007956986	& &	0.8042036141630704	& &	0.8092189699904901	\cr
 3 & & 0.8054651813042190	& &	0.8096942202109880	& &	0.8159975210361090	\cr
 4 & & 0.8037461158266676	& &	0.8035523554262308	& &	0.8036525961629886	\cr
 5 & & 0.8038036096317446	& &	0.8040771905224384	& &	0.8048413909234290	\cr
 6 & & 0.8037708004664887	& &	0.8037491193175370	& &	0.8037061119915903	\cr
 7 & & 0.8037713901307034	& &	0.8037870364646635	& &	0.8038694389755376	\cr
 8 & & 0.8037706721375962	& &	0.8037690121024319	& &	0.8037599392046063	\cr
 9 & & 0.8037706545007351	& &	0.8037715434308201	& &	0.8037800051832189	\cr
 10 & & 0.8037706342530531	& &	0.8037705356047065	& &	0.8037692091978080	\cr
 11 & & 0.8037706322885981	& &	0.8037706977772240	& &	0.8037715412805357	\cr
 12 & & 0.8037706314198754	& &	0.8037706406762437	& &	0.8037704590622494	\cr
 13 & & 0.8037706312900128	& &	0.8037706508130255	& &	0.8037707215178032	\cr
 14 & & 0.8037706312192222	& &	0.8037706475443096	& &	0.8037706116069788	\cr
 15 & & 0.8037706312169512	& &	0.8037706481678014	& &	0.8037706405540735	\cr
\hline
\hline
\end{tabular}
\caption{ Convergent value for ground state energy $E_0$ of an AHO with $\lambda =1$ in  MFNPS using multi step optimal splitting parameters.}
\label{tab2}
\end{table}

In this section we introduce a multi step optimal splitting approach on parameter calculation
with a aim to improve the convergency.
Following the previous procedure, we write the expression for $H_{D}$ as  
\begin{equation}
H_{D}=\frac{\left(W_k + \frac{\textstyle{1}}{\textstyle{W_k}}\right) (2\,a^{\dagger}a +1)}{4} +
\lambda \, \,\frac{3+ 12 a^{\dagger}a + 6 (a^{\dagger})^2 a^2}{4W_k^2} 
\label{eqn20}
\end{equation}

Similarly the expression for $H_{N}$ as
\begin{eqnarray}
H_{N} &=& \lambda [ \frac{a^4 + (a^{\dagger})^4 + 4 (a^{\dagger})^3 a + 4 a^{\dagger} a^3}{4W_k^2} \nonumber \\
&-& \frac{3(a^2 + (a^{\dagger})^2)}{2  W_k} \left( \frac{1}{W_{k-1}} + \frac{1}{W_{k-2}} + \cdots + 
\frac{1}{W_{1}} + \frac{1}{w} \right ) ]
\label{eqn21}
\end{eqnarray}

where $k$ is the order of multi step optimal splitting. 
The optimal splitting parameters $W_{k}$ are given by
\begin{equation}
W_{k}^{2}= W_{k-1}^{2} + \frac {6\lambda}{W_{k}}
\label{eqn22}
\end{equation}
with 
\begin{equation}
W_{1}^{2}= w^{2} + \frac {6\lambda}{W_{1}}
\label{eqn23}
\end{equation}

The non zero expectation values of diagonal Hamiltonian $H_D$ are given as 

\begin{equation}
\langle n |H_D| n \rangle =\left (\frac{2n+1}{4}\right) \left(W_k + \frac{1}{W_k}\right) + \frac{3 \lambda}{4 W_k^2}(2n^2 + 2n +1)
\label{eqn24}
\end{equation}
The non zero expectation values of non diagonal Hamiltonian $H_N$ are given as
\begin{equation}
\langle n |H_N| n+2 \rangle =\frac{ \lambda \sqrt{(n+1)(n+2)}}{W_k} \, \left[\frac{n}{W_k} - \frac{3}{2}
\left (\frac{1}{W_{k-1}} + \frac{1}{W_{k-2}}+ \cdots + \frac{1}{W_1}+\frac{1}{w}\right) \right]
\label{eqn25}
\end{equation}

\begin{equation}
\langle n |H_N| n+4 \rangle =\frac{ \lambda \sqrt{(n+1)(n+2)(n+3)(n+4)}}{4W_k^2}
\label{eqn26}
\end{equation}

Here considering different values of $k$, we calculate the groundstate energy 
for $\lambda=1$ and the results are given in Table \ref{tab2}. We notice that 
two step optimal splitting is sufficient to provide convergency in MFNPS for 
ground state of AHO.

Two step optimal splitting approach is also applied to the ground state of AHO for higher coupling parameter 
$\lambda$. Last three higher order convergent results for different coupling parameter $\lambda$ is given in 
Table \ref{tab3}

\begin{table}[h]
\centering
\begin{tabular}{cccccccc}
\hline
\hline
  	&  &   Convergent value     & &  Convergent value      & &  Convergent value     \cr
 Order  &  &   $E_0$ for            & &  $E_0$ for             & &  $E_0$ for            \cr
 K      &  &   $\lambda=0.01$       & &  $\lambda=10$          & &  $\lambda=100$      \cr

\hline
\hline
 13 & & 0.5072562106523008& &	1.504972427622453& &	3.131384278221808\cr
 14 & & 0.5072562106523008& &	1.504972406538491& &	3.131384221926994\cr
 15 & & 0.5072562106523008& &	1.504972417516847& &	3.131384248591904\cr
\hline
\hline
\end{tabular}
\caption{ Convergent value for ground state $E_0$ of an AHO with different values of coupling parameter  $\lambda$ in  MFNPS using two-step optimal splitting. 
parameters.}
\label{tab3}
\end{table}

\section{State-independent parameter calculation on excited state using two step optimal splitting}
In this section we apply previously calculated optimal splitting parameters for ground state of AHO
to the exited state of AHO. The results are given in Table \ref{tab5}. We observe that for 
$n \ge 5$ the MFNPS is not giving any convergent value and also for $n \le 5$ the number of convergent digits 
decreases as the increase of state value $n$. Again the convergent results are not comparable for higher 
excited states with the results calculated previously by several authors [2-4]. This is expected as the 
excited state energy depends on the state value $n$. So the optimal splitting parameters has to be modified 
according to their state dependency. 

%

\begin{table}[h]
\centering
\begin{tabular}{lcccc}
\hline
\hline
	     & $E_n$ for     & $E_n$ for       &  $E_n$ for      \cr
State	     & Coupling      & Coupling        & Coupling       \cr
 $E_n$       & parameter     &  parameter      &  parameter     \cr
	     & $\lambda=0.1$ &  $\lambda=1$    &  $\lambda=100$ \cr

\hline
\hline
$1$&1.769502633601580 & 2.737893473247960	 & 11.18727013754662	 	\cr
$2$&3.138624640483820 & 5.179368610682413	 & 21.90792389514790	 	\cr
$3$&4.628893580258386 & 7.944276200342911	 & 34.20660264641097	 	\cr
$4$&6.220490587163873 & 10.98830903940391	 & 48.00716948459711 	 	\cr
$5$&7.901913609979696 & 14.41671095662173 	 & 65.17919098201246	 	\cr
$6$&9.674274270406038 & NC	 & NC	 	\cr
$7$&11.58029965657092 & NC	 & NC	 	\cr
$8$&NC 				    & NC	 & NC	 	\cr
$9$&NC  			    & NC	 & NC	 	\cr
$10$&NC  			    & NC	 & NC	 	\cr
\hline
\hline
\end{tabular}
\caption{ Convergent value for excited state energy $E_n$ of an AHO using $15$ order of terms of the MFNPS}
\label{tab5}
\end{table}
NC-{\it No Convergency}
\section{State-dependent parameter calculation on excited state using single step parameter}
The state dependent parameter $W$ is determined in such a way that it will make $H_D$ and
$H_N$ as small as possible simultaneously. One of the simplest ways is to determine $W$ from $H_D$ for the
desired state, say $n$, using the variational principle
\begin{equation}
 \frac{d \langle n |H_D| n \rangle}{dW} =0
\end{equation}

$W$ is determined from the cubic equation 
\begin{equation}
W^{3}-  W  - \frac {6\lambda(2 n^{2}+2 n +1)}{(2 n +1 )}=0
\label{eqn27}
\end{equation}

For ground state $n=0$ the above equation reduces to the state independent equation as discussed in 
Eq \ref{eqn10}.

\begin{table}[h]
\centering
\begin{tabular}{ccccc}
\hline
\hline
	     & $E_n$ for     & $E_n$ for      &  $E_n$ for      \cr
State        & Coupling      & Coupling        & Coupling       \cr
$n$          & parameter     &  parameter      &  parameter     \cr
             & $\lambda=0.1$ &  $\lambda=1$    &  $\lambda=100$ \cr

\hline
\hline
$1$  & 1.769502526042401  & 2.737826568159874  	 & 11.18590665985774  	\cr
$2$  & 3.138624260783390  & 5.179278503183649 	 & 21.90667970339865	\cr
$3$  & 4.628882799032086  & 7.942400664976592 	 & 34.18247662808967	\cr
$4$  & 6.220300899387960  & 10.96358201989640 	 & 47.70719233011905	\cr
$5$  & 7.899767255018711  & 14.20313874402137  	 & 62.28123351170720	\cr
$6$  & 9.657840024381196  & 17.63404895868295 	 & 77.77076877553597	\cr
$7$  & 11.48731562245528  & 21.23643543180502 	 & 94.07804796313151	\cr
$8$  & 13.38247490877576  & 24.99493650263666 	 & 111.1279601151626	\cr
$9$  & 15.33864203774306  & 28.89725106727866  	 & 128.8606294268705 	\cr
$10$&  17.35190767828287  & 32.93326317015690 	 & 147.2269956471010	\cr
\hline
\hline
\end{tabular}
\caption{ Energy levels $E_n$ of the AHO using $K=14$ order of terms in MFNPS using 
single step variational parameters.}
\label{tab6}
\end{table}
Now with this variational parameter the non zero expectation value of diagonal Hamiltonian $H_D$ is given as 

\begin{equation}
\langle n |H_D| n \rangle =\frac{2n+1}{4}\left(W + \frac{1}{W}\right) + \frac{3 \lambda}{4 W^2}(2n^2 + 2n +1)
\label{eqn28}
\end{equation}
The non zero expectation value of non diagonal Hamiltonian $H_N$ is given as 
\begin{equation}
\langle n |H_N| n+2 \rangle = \left[\frac{1}{4}\left(-W+\frac{1}{W}\right) + \frac{\lambda}{W^2}\left(n+ \frac{3}{2}\right)\right] \sqrt{(n+1)(n+2)}
\label{eqn29}
\end{equation}

\begin{equation}
\langle n |H_N| n+4 \rangle =\frac{ \lambda \sqrt{(n+1)(n+2)(n+3)(n+4)}}{4W^2}
\label{eqn30}
\end{equation}

We calculate the excited state energies of an AHO up to $14$ order and $15$ order terms in MFNPS using 
single step variational parameter and the results are given in Table \ref{tab6} and Table \ref{tab7}
respectively. Camparing results from Table \ref{tab6} and Table \ref{tab7}, we achieve a convergency 
up to a minimum of $7$ digit. 

\begin{table}[h]
\centering
\begin{tabular}{ccccc}
\hline
\hline
	     & $E_n$ for     & $E_n$ for      &  $E_n$ for      \cr
State        & Coupling      & Coupling        & Coupling       \cr
$n$          & parameter     &  parameter      &  parameter     \cr
             & $\lambda=0.1$ &  $\lambda=1$    &  $\lambda=100$ \cr

\hline
\hline
$1$  & 1.769502734911583  & 2.737955961049832  	 &  11.18865883835119  	\cr
$2$  & 3.138624351464486  & 5.179303188325927 	 &  21.90710180727473	\cr
$3$  & 4.628882837847637  & 7.942406660726320 	 &  34.18256466416197	\cr
$4$  & 6.220300917364425  & 10.96358385150704 	 &  47.70721635883445	\cr
$5$  & 7.899767264110872  & 14.20313941758235  	 &  62.28124490910177	\cr
$6$  & 9.657840029418940  & 17.63404924938320 	 &  77.77077204538597	\cr
$7$  & 11.48731562551444  & 21.23643557566123 	 &  94.07804949179396	\cr
$8$  & 13.38247489117164  & 24.99493657142581 	 &  111.1279607854576	\cr
$9$  & 15.33864207925654  & 28.89725105542083	 &  128.8606292267516	\cr
$10$ & 17.35190767499986  & 32.93326304139077 	 &  147.2269943696901	\cr
\hline
\hline
\end{tabular}
\caption{ Energy levels $E_n$ of the AHO using $K=15$ order of terms in MFNPS using 
single step variational parameters.}
\label{tab7}
\end{table}

\section{State-dependent parameter calculation on excited state using two step parameter}

\begin{table}[h]
\centering
\begin{tabular}{ccccc}
\hline
\hline
	     & $E_n$ for     & $E_n$ for      &  $E_n$ for      \cr
State        & Coupling      & Coupling        & Coupling       \cr
$n$          & parameter     &  parameter      &  parameter     \cr
             & $\lambda=0.1$ &  $\lambda=1$    &  $\lambda=100$ \cr

\hline
\hline
$1$  & 1.769502595495307  &  2.737892280858456 	 &  11.18829020411844  	\cr
$2$  & 3.138624197794214  &  5.179291722504740	 &  21.90689767968495	\cr
$3$  & 4.628882511628982  &  7.942403919498243	 &  34.18252348780879	\cr
$4$  & 6.220300863131970  &  10.96358293850466	 &  47.70720519550517	\cr
$5$  & 7.899767113294260  &  14.20313925236565 	 &  62.28123822043900 	\cr
$6$  & 9.657840059798382  &  17.63404889881106	 &  77.77077201647030	\cr
$7$  & 11.48731530776505  &  21.23643596676358	 &  94.07805770355252	\cr
$8$  & 13.38247452708933  &  24.99493915652584	 &  111.1279999324749	\cr
$9$  & 15.33864161377721  &  28.89726023457015 	 &  128.8606564368714	\cr
$10$&  17.35190770598516  &  32.93326418012689	 &  147.2269846271021	\cr
\hline
\hline
\end{tabular}
\caption{ Energy levels $E_n$ of the AHO using $K=14$ order of terms in the MFNPS perturbation 
series using two step optimal splitting variational parameters.}
\label{tab8}
\end{table}
In this section we introduce state dependent parameter using two step optimal splitting 
procedure for the convergency test on excited states.
The two optimal splitting parameters $W$ and $w$ are calculated from the cubic equation given by
\begin{equation}
W^{3}- w^{2} W  - \frac {6\lambda(2 n^{2}+2 n +1)}{(2 n +1 )}=0
\label{eqn31}
\end{equation}
where 
\begin{equation}
w^{2}= 1 + \frac {6\lambda}{w}
\label{eqn32}
\end{equation}

The non zero expectation value diagonal Hamiltonian $H_D$ is given as 

\begin{equation}
\langle n |H_D| n \rangle =\left(\frac{2n+1}{4}\right)\left(W + \frac{1}{W}\right) + \frac{3 \lambda}{4 W^2}(2n^2 + 2n +1) 
\label{eqn33}
\end{equation}

\begin{equation}
\langle n |H_N| n+2 \rangle = \sqrt{(n+1)(n+2)} \, \left[\frac{-W}{4} + \frac{1}{4 W} + \frac{\lambda}{W^2}
\left (n +\frac{3}{2}\right) \right ] 
\label{eqn34}
\end{equation}

The non zero expectation value non diagonal Hamiltonian $H_N$ is given as 

\begin{equation}
\langle n |H_N| n+4 \rangle =\frac{ \lambda \sqrt{(n+1)(n+2)(n+3)(n+4)}}{4W^2}
\label{eqn35}
\end{equation}

The exicited energy levels calculated up to $14$ and $15$ order of terms in MFNPS using 
two step optimal splitting variational parameters and the resutls are given in 
Table \ref{tab8} and Table \ref{tab9}

\begin{table}[h]
\centering
\begin{tabular}{ccccc}
\hline
\hline
	     & $E_n$ for     & $E_n$ for      &  $E_n$ for      \cr
State        & Coupling      & Coupling        & Coupling       \cr
 $n$         & parameter     &  parameter      &  parameter     \cr
             & $\lambda=0.1$ &  $\lambda=1$    &  $\lambda=100$ \cr

\hline
\hline
$1$ & 1.769502595720013  & 2.737892290808430	 & 11.18725175268262 	\cr
$2$ & 3.138624197891925  & 5.179291724439308	 & 21.90689768817313	\cr
$3$ & 4.628882511666889  & 7.942403920161726	 & 34.18252348715269 	\cr
$4$ & 6.220300863145563  & 10.96358293822675	 & 47.70720518995203 	\cr
$5$ & 7.899767113290646  & 14.20313925025467	 & 62.28123817953492	\cr
$6$ & 9.657840059675530  & 17.63404886347468	 & 77.77077131768290	\cr
$7$ & 11.48731530570292  & 21.23643553577421	 & 94.07805012397570 	\cr
$8$ & 13.38247450558015  & 24.99493599000598 	 & 111.1279501536521	\cr
$9$ & 15.33864146634275  & 28.89724877539235	 & 128.8606557529643	\cr
$10$& 17.35190696572501  & 32.93326503128077	 & 147.2269836271021	\cr
\hline
\hline
\end{tabular}
\caption{ Energy levels $E_n$ of the AHO using $K=15$ order of terms in the MFNPS perturbation 
series using two step optimal splitting variational parameters.  }
\label{tab9}
\end{table}

\section{Conclusion}
In this paper we have taken a very fruitful approach to get convergency in MFNPS in order to 
get energy levels of an AHO. The problem of energy levels of AHO has been analysed from wide 
angles to achieve convergency in MFNPS. MFNPS seems to be one of most simplest non linear 
series to be used for energy levels of AHO. Optimal splitting method is found to be very 
efficient technique to achieve convergency in MFNPS.

\newpage

\bigskip

\begin{spacing}{1.5}
\begin{small}

\end{small}
\end{spacing}

\end{document}